\documentclass[12pt,a4paper]{article}
\usepackage{epsfig}

\begin{document}

\begin{titlepage}

\begin{flushright}
  MC-TH-2002-4  \\ 
  hep-ph/0207283
\end{flushright}
 
   \begin{center}

    \vskip 20mm
    {\Large\bf An eikonal model for multiparticle production in hadron-hadron interactions}
    \vskip 5mm
   
    {\large I.Borozan$^{1,2}$ and M.H.Seymour$^{3}$}\\[0.25cm]
    {\it $^1$Rutherford Appleton Laboratory, Chilton, Didcot, Oxon. OX11 0QX, U.K.}\\
    {\it $^2$University College London, Gower Street, London. WC1E 6BT, U.K.}\\
    {\it $^3$University of Manchester, Oxford Rd., Manchester, M13 9PL, U.K.}\\

  \end{center}
  \vskip 1mm
  \begin{abstract}
    \noindent

We introduce an eikonal Monte Carlo model (running in conjunction with HERWIG) for simulating multiparticle production in hadron-hadron interactions. We compare our simulated data to the CDF Tevatron measurement of the underlying event activity in hard inelastic  proton-antiproton scattering at $\sqrt{s}$=1.8 TeV. By fixing the only free parameter in our model, the total hadron-hadron cross section, we find that our model describes the data better than either the HERWIG Underlying Event model or the Hard Multiparton model.

  \end{abstract}

\end{titlepage} 

\section{Introduction}

In this paper we introduce an eikonal Monte Carlo model which simulates hadron-hadron collisions in which a ``hard'' parton-parton scattering has occurred. In hard events, described by perturbative QCD, the outgoing particles with large transverse momentum form jets. These are clusters of outgoing hadrons produced approximately in the direction of the two outgoing partons of the hard parton-parton scattering. Jet physics plays an essential role as a test of perturbative QCD and in determining  the parameters of the Standard Model. It will be important for future detection of new particles and might help shed light on the nature the Pomeron. 

If we are to find some new physics at future (or current colliders) or test the known physics (such as perturbative QCD), we will need to give an accurate description of the already known physical process which would then be considered as a background to the new physics. 

The underlying event, in hard interactions, is defined to be all the additional interactions that are not part of the hard scattering. Thus the underlying event may contain particles from proton-antiproton, remnant-remnant interactions and hard multiparton interactions, if such interactions do occur. 

At present, our understanding of underlying event physics is not sufficient to match the precision of experimental measurements, so to fully exploit current and future data we need to improve that understanding.

To this end the CDF collaboration recently measured the activity of the underlying event in proton-antiproton hard scattering \cite{Affolder:2002zg}. They found that its activity is considerably larger than in soft collisions and that none of the QCD Monte-Carlo models describe correctly  all the properties of the underlying event (however the best results were obtained with Pythia \cite{Sjostrand:1987su}). 

Our goal is to introduce a simple eikonal Monte Carlo model, with a small number of parameters, and compare it to the measured data from the CDF collaboration \cite{Affolder:2002zg}. In Sec.2 and Sec.3 we present the overall theoretical framework of the model and give a description of the hard part, in Sec.4 we present our assumptions behind the soft subprocess and its implementation within the HERWIG Monte Carlo package and in Sec.5 we compare our theoretical results with those from the CDF experiment \cite{Affolder:2002zg} and two other underlying event models, the HERWIG underlying event model \cite{Marchesini:1992ch} and the Hard Multiparton model \cite{Butterworth:1996zw}.

\section{Optical theorem and impact parameter representation}
In this section we will review some well known expressions for cross sections within the eikonal approximation, which we will use as the basic framework for our model.

For two incoming particles $\it{A, B}$ and two outgoing particles  $\it{C, D}$ we write their four-momenta as $p_{A}, p_{B}, p_{C}, p_{D}$ and define the invariant Mandelstam variables $\it{s}$ and $\it{t}$ as:

\begin{equation}
s = (p_{A} + p_{B})^{2}, \; t = (p_{A} - p_{C})^{2}.
\end{equation}

The optical theorem relates the imaginary part of the elastic scattering amplitude $A(s,t=0)$ to the total cross section $\sigma_{tot}$ through a unitarity relation \cite{collins:Intro}:

\begin{equation}    
\sigma_{tot}=\frac{1}{s}Im(A(s,t=0)). 
\end{equation}
Neglecting effects that depend on the spin of the scattering particles we can express the scattering amplitude as the Fourier transform of the elastic scattering amplitude in impact parameter ${\it{a({\bf{b}}, s)}}$ (${\bf{b}}$ being the impact parameter and ${\bf{q}}$ being the transverse component of momentum transfer in the CMS) as :

\begin{equation}    
A(s,t) = 4 \pi s \int db^2 a( {\bf{b}}, s) e^{i \bf{q.b}}. 
\end{equation}
Using the optical theorem (2) we can express the cross sections as:
\begin{equation}    
\sigma_{tot} = 4 \pi \int db^{2} Im(a({\bf{b}},s)), 
\end{equation}   
\begin{equation}    
\sigma_{ela} = 4 \pi \int db^2 |a({\bf{b}},s)|^{2}, 
\end{equation}
\begin{equation}    
\sigma_{inel} = \sigma_{tot}-\sigma_{ela}.                       
\end{equation}
We express the elastic amplitude in terms of the eikonal function, $\chi(b,s)$, as \cite{collins:Intro}:
\begin{equation}    
a({\bf{b}},s) = \frac{e^{-\chi({\bf{b}},s)}-1}{2i}. 
\end{equation}
With the eikonal expression (7) we can write the cross sections as :
\begin{equation}    
\sigma_{tot} = 2 \pi \int_{0}^{\infty} db^{2} [1-e^{-\chi(b,s)}], 
\end{equation}
\begin{equation}    
\sigma_{ela} = \pi \int_{0}^{\infty} db^{2} \left |[1-e^{-\chi(b,s)}] \right |^{2}, 
\end{equation}
\begin{equation}    
\sigma_{inel} = \pi \int_{0}^{\infty} db^{2} [1-e^{-2\chi(b,s)}]. 
\end{equation}
We choose the total cross section in (8) as the main parameter of our model. In the next section we will specify the expression for ${\chi(b,s)}$.

\section{Expression for the eikonal}
In the following section, we will give an expression for the eikonal ${\chi(b,s)}$ in (8) and provide some justification for its expression.

\subsection{The hard part of the eikonal} 
We start by assuming that at high center of mass energy ($\sqrt{s}$ $>$ 100 GeV), the main hard scattering is accompanied by semihard interactions, which arise as a result of hard multiple parton scattering in which each parton carries a very small fraction of its parent hadron's momentum. Multiparton interactions lead to the appearance of minijets, i.e. jets with transverse energy much smaller than the total energy ($\sqrt{s}$) available in the hadron-hadron collision. Multiparton interactions will then drive the rise of the inclusive jet cross section at high energies depending on the small-$\it{x}$ behaviour of the parton distribution or, to a good approximation, that of the dominant gluon distribution.

In refs.[\cite{Butterworth:1996zw} \cite{Collins:Ladyn}, \cite{Forshaw:Stor}, \cite{Fletch:andall}, \cite{Durand:Pi}] it was shown that the average number of secondary hard scatters in an event can be deduced from the QCD improved parton model. For our hard multiple-interactions we follow the model presented in \cite{Butterworth:1996zw} where the average number of secondary hard scatters $\langle\it{n(b,s_{p\overline{p}})}\rangle$ is given by :
\begin{equation}
\langle\it{n(b,s_{p\overline{p}})}\rangle = \it{A(b)\sigma_{H}^{inc}(s_{p\overline{p}})},
\end{equation} 
where the profile function $\it{A(b)}$ specifies the overlap of partons in the two hadrons in impact parameter. It can be written as a convolution of form factor distributions of two incoming hadrons, 
\begin{equation}
A(b) = \int d^{2}\textbf{b'}G_{p}(\textbf{b'})G_{\overline{p}}(\textbf{b}-\textbf{b'}),
\end{equation}
for which we take
\begin{equation}
G_{p}(\textbf{b}) = G_{\overline{p}}(\textbf{b})=\int \frac{d^{2}\textbf{k}}{(2\pi)}\frac{\exp{(\textbf{k $\cdot$ b})}}{(1+\textbf{k}^{2}/ \mu^{2})^{2}},
\end{equation}
with $\mu^{2} = 0.71$ GeV$^2$.
The integral then yields \cite{Fletch:andall}: 
\begin{equation}
A(b) = \frac{\mu^{2}}{96\pi}(\mu b)^{3}K_{3}(\mu b),
\end{equation}
with $K_{i}(x)$ the modified Bessel function. The overlap function $\it{A(b)}$ satisfies
\begin{equation}
\int \pi db^2A(b) = 1.
\end{equation}
The cross section appearing in (11), $\it{\sigma_{H}^{inc}(s_{p\overline{p}})}$, is the inclusive cross section for $\it{\overline{p}p}$$\rightarrow$ jets with $\it{p_t}$$ > $$\it{p_{tmin}}$ :
\begin{eqnarray}
\sigma_H^{inc}(s_{p\overline{p}}) & = & \int^{s_{p\overline{p}}/4}_{p_{tmin}^2}dp_{t}^2\int^1_{4p_t^2/s_{p\overline{p}}}dx_p\int^1_{4p_t^2/x_{p}s_{p\overline{p}}}dx_{\overline{p}}\sum_{i,j}f_i(x_p,\ p_t^2)f_j(x_{\overline{p}},\ p_t^2)\nonumber \\
                                  &   & \times\frac{d\hat{\sigma_{ij}}(x_px_{\overline{p}}s_{p\overline{p}},p_t)}{dp_t^2}.
\end{eqnarray}
It is assumed, (as in ref.[\cite{Sjostrand:1987su},\cite{Butterworth:1996zw}]) that secondary hard scatters are independent of each other so that the probability for the number of hard scatters $\it{h}$, in a ${p\overline{p}}$ collision at a given value of impact parameter, is given by the Poissonian probability distribution,   
\begin{equation}
P_h=\frac{(\langle\it{n(b,s_{p\overline{p}})\rangle)}^h}{h!}\exp{(-\langle\it{n(b,s_{p\overline{p}})\rangle)}}.
\end{equation}       
The inelastic hard cross section for $\it{p\overline{p}}$$\rightarrow$partons with $\it{p_t}$$ > $$\it{p_{tmin}}$ can be written as :
\begin{eqnarray}
 \sigma_H(s_{p\overline{p}}) & = & \pi\int{d{b^2}}\sum_{h=1}^{\infty}P_h\nonumber \\
                           & = & \pi\int{d{b^2}}[1-\exp(-{\langle}n(b,s_{p\overline{p}})\rangle)].              \end{eqnarray}

The hard  multiparton part \cite{Butterworth:1996zw} of the model was already implemented as part of the HERWIG Monte Carlo program \cite{Marchesini:1992ch}. Here we give a brief summary.  

Event simulation starts with HERWIG generating a hard subprocess according to the leading order cross section. Both incoming and outgoing partons involved in the hard subprocess (at some hard scale $\it{Q}$) are evolved (backward for the incoming parton) through a coherent parton shower algorithm until they reach a typical hadronic scale, $Q_{h}$ $\approx$ 1GeV. The coherent parton shower algorithm \cite{herw:shower} resums to all orders both single logarithmic terms associated with the collinear emission (through implementation of the DGLAP splitting function) and single and double logarithmic terms associated with soft emissions. All the successive partons emitted in the parton shower are colour connected in such a way that partons carrying colour(anticolour) tend to end up close in momentum and real-space to their anticolour(colour) partners. Once the evolved partons have reached a hadronisation scale $Q_{h}$, clusters of partons are formed where each colour connected pair of partons forms a single colourless cluster. Each cluster is then decayed into hadrons according to phase space arguments \cite{Webber:1983if}. Once the main hard subprocess has been simulated, energy-momentum conservation is used to calculate the momentum of the remaining diquark(antidiquark) or proton(antiproton) remnant. Each diquark carries opposite colour to that of the parton involved in the hard subrocess. After the first hard scattering has been completed a number of secondary hard subprocess are simulated where each coloured remnant from the previous interaction is labeled as the new incoming hadron from which the quark valence distribution functions have been removed.

Comparing the eikonal expression in (10) with the cross section in (18), we define the hard part of the eikonal as :

\begin{equation}
\chi_{QCD}(b,s) = \frac{1}{2}\langle\it{n(b,s_{p\overline{p}})}\rangle.
\end{equation}
\noindent 

The QCD perturbative 2-to-2 parton-parton differential cross section diverges as the transverse momentum of the scattering, $\it{p_{t}}$, goes to zero. One must then fix a minimum value $\it{p_{tmin}}$ for $\it{p_{t}}$ large enough so that the resulting cross section is not larger than the total inelastic non-diffractive cross section and also large enough for perturbative QCD to be reliable. This dependence of the perturbative QCD differential cross section implies that the eikonalised inelastic hard cross section would also depend on the arbitrary value chosen for the minimum $\it{p_{t}}$. 

\subsection{The expression for ${\chi_{total}(b,s)}$}
Our goal is to formulate a model which will be to a certain degree independent of $\it{p_{tmin}}$ cutoff.
To this end, to $\chi_{QCD}(b,s_{p\overline{p}})$ which describes hard interactions with $\it{p_t}$$ \geq $$\it{p_{tmin}}$, we will add $\chi_{soft}(b,s_{p\overline{p}})$, which will describe interactions with 0 $\leq$ $\it{p_t}$$ \leq $$\it{p_{tmin}}$. The full eikonal can now be expressed as :

\begin{equation}
\chi_{total}(b,s) = \chi_{QCD}(b,s_{p\overline{p}}) + \chi_{soft}(b,s_{p\overline{p}}). 
\end{equation} 

\noindent
To specify $\chi_{soft}(b,s_{p\overline{p}})$ we will use a model introduced by Chou and Yang \cite{Chou:Yang} and Durand and Pi \cite{Durand:Pi}, which postulates that the elastic scattering is the shadow of the absorption resulting from the passage of one hadronic mass distribution through another. The transverse distribution of the matter is assumed then to have the same shape as the charge distribution, as measured by the electromagnetic form factor, so that $\chi_{soft}(b,s)$ is of the form:

\begin{equation}
\chi_{soft} = C(s_{p\overline{p}})A(b),
\end{equation} 
where ${C(s_{p\overline{p}})}$ is a constant to be determined below and ${A(b)}$ is given in (13).

The eikonal now consists of two parts, the hard, 
\begin{equation}
\chi_{QCD} = \frac{1}{2}\sigma_H^{inc}(s_{p\overline{p}})A(b),
\end{equation}
and the soft,
\begin{equation}
\chi_{soft} = \frac{1}{2} \sigma_{SOFT}^{inc}(s_{p\overline{p}}) A(b).
\end{equation}
If we now consider the total cross section (8) with $\chi(b,s)$=$\chi_{total}(b,s)$ and assume $\chi_{total}(b,s)$ to be small, we can expand the exponential in (8). Integrating over the impact parameter $\it{b}$ we get:

\begin{equation}  
\sigma_{total}(s_{p\overline{p}}) \approx \sigma_{SOFT}^{inc}(s_{p\overline{p}}) + \sigma_H^{inc}(s_{p\overline{p}})
\end{equation}

Following this approximation we will assume that our $\sigma_{SOFT}^{inc}(s_{p\overline{p}})$ is a bare non-perturbative cross section for soft proton-antiproton interaction. Since $\sigma_{SOFT}^{inc}(s_{p\overline{p}})$, the bare soft cross section in (23), is not directly calculable in our model, its value can be determined from the experimental data, i.e. by using the total cross section measured by the CDF collaboration \cite{tot:ppbar}. 

To give a prediction for future LHC data we will rely on the Donnachie-Landshoff model of the Pomeron \cite{Donnachie:1992ny} which fits successfully all the experimental data on total cross section for proton-proton scattering, 

\begin{equation}
\sigma_{tot}(s_{pp}) = \sigma_{tot}(s_{0}) \left(\frac{s}{s_{0}}\right)^{\alpha_{p}(0)-1},
\end{equation} 
where ${\alpha_{p}(t) = 1+\epsilon+\alpha_{p}^{'}t}$ with ${\epsilon = 0.08}$ and ${\alpha_{p}^{'} = 0.25}$ GeV$^{-2}$.  
Assuming ${\sigma_{tot}(s_{pp})}$ = ${\sigma_{tot}(s_{p\overline{p}})}$ at high energy, the total cross section at the LHC can then be written as 
\begin{equation}
\sigma_{tot}(s_{pp}) = \sigma_{tot}(s_{CDF})\left(\frac{\sqrt{s_{pp}}}{1800}\right)^{2 \times 0.08}.
\end{equation}

We consider the value of the total cross section as the only phenomenological input to our model (in addition to those that are already in HERWIG\cite{Marchesini:1992ch} and the multiparton hard model \cite{Butterworth:1996zw}, which we leave at their default values). \\

In addition the model contains the following assumptions : 

\vskip 2mm
$\bullet$ there can be more than one soft interaction in a soft proton-antiproton interaction. \\


$\bullet$ the probability distribution for having $\it{m}$ and only $\it{m}$ soft scatters in a given proton-antiproton interaction $P_m$ obeys the Poissonian distribution (i.e. multiple soft scatters are uncorrelated), 

\begin{equation}
P_m = \frac{(2\chi_{soft})^m}{m!}\exp{(-2\chi_{soft})}.
\end{equation}

$\bullet$ the hard and soft scatters are assumed to be independent i.e the probability distribution for having $\it{h}$ hard, and $\it{m}$ soft, scatters is given by $P_{h,m}$:

\begin{equation}
P_{h,m}(b,s_{p\overline{p}}) = \frac{ (2 \chi_{QCD})^h}{h!}\frac{(2 \chi_{soft})^m}{m!}\exp{(-2 \chi_{total})}.
\end{equation}

\vskip 4mm

We can now express cross sections as:

\begin{equation}
\sigma_{inel}(s_{p\overline{p}}) = \pi \int db^{2} \sum_{h+m \geq 1} P_{h,m}(b,s_{p\overline{p}}) = \pi\int_{0}^{\infty}db^2(1-e^{-2\chi_{total}(b,s_{p\overline{p}})}),
\end{equation}
the inelastic cross section with at least one hard scattering:
\begin{equation}
\sigma_{h,inel}(s_{p\overline{p}}) = \pi \int db^{2} \sum_{h \geq 1, m \ge 0} P_{h,m}(b,s_{p\overline{p}}) = \pi\int_{0}^{\infty}db^2(1-e^{-2\chi_{QCD}(b,s_{p\overline{p}})}),
\end{equation}
and the total cross section:
\begin{equation}
\sigma_{tot}(s_{p\overline{p}}) = 2\pi\int_{0}^{\infty}db^2(1-e^{-\chi_{total}(b,s_{p\overline{p}})}).
\end{equation}


The $\sigma_{SOFT}^{inc}(s_{p \overline{p}})$ in (23) is then determined by fixing its value so that the the total cross section (31), at some $p_{tmin}$, is equal to the total cross section measured by the CDF collaboration ($\sigma_{total\;CDF}$ = 81.8$\pm$2.3mb) (see Sec.5 for different values of $\sigma_{SOFT}^{inc}(s_{p \overline{p}})$ and $p_{tmin}$ used).

In order to have a realistic Monte Carlo event generator we have implemented our eikonal model as part of the HERWIG Monte Carlo program \cite{Marchesini:1992ch} with the full use of its hadronisation model and parton showering properties.

\section{The Monte Carlo implementation of hard and soft processes}
In the following sections we will describe soft processes in detail. The hard processes are as in \cite{Butterworth:1996zw} (see Sec.3).
Before we start with the simulation of an event we need to fix $\sigma_{SOFT}^{inc}(s_{p \overline{p}})$, the bare soft cross section (as explained in Sec.3).  Then by drawing from the probability distribution (28) we decide the number of soft and  hard interactions in a given event. The event simulation can then start with HERWIG generating a hard subprocess followed by multiparton ones (see Sec.3). Once the multiple hard scatters have been exhausted the soft multiple scatters are generated.

\subsection{Assumptions behind the soft subprocess}

At high $\it{s_{p\overline{p}}}$ (center of mass energy) and low $Q$ scale of interaction, i.e low-$\it{x}$, we might expect a proliferation of soft gluons.
In order to simulate each soft subprocess we assume that the bare soft cross section $\sigma_{SOFT}^{inc}(s_{p\overline{p}})$, corresponds to a soft interaction between two incoming soft gluons. That is, we model the soft inelastic collision between two remnants as a soft elastic collision between two partons within them (see fig.1).




All soft gluons carry a colour charge and have an effective mass $m_{g}$ (inspired by HERWIG's model of the hadronisation of the outgoing gluons). 
The two outgoing effective gluons in fig.1 are on mass shell and colour connected to the remnants and to each other, as shown in fig.1, each gluon colour(anticolour) is connected to its anticolour(colour) partner.

\begin{figure}[h]
\begin{center}
\epsfig{figure=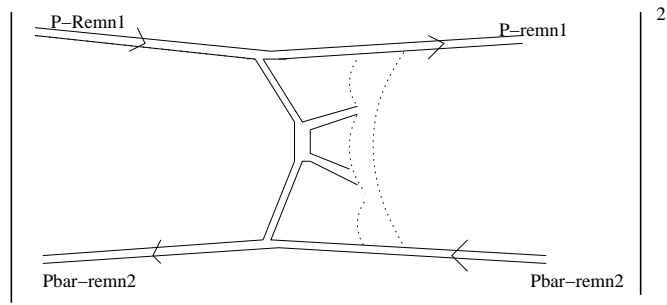,height=4cm}
\end{center}\caption{\footnotesize $\sigma_{SOFT}^{inc}(s_{p\overline{p}})$ corresponds to a soft collision between the two soft gluons (full color picture). Remnants are also connected to each other via $t$ channel gluon line.}
\end{figure}
\noindent

\begin{figure}[h]
\begin{center}
\epsfig{figure=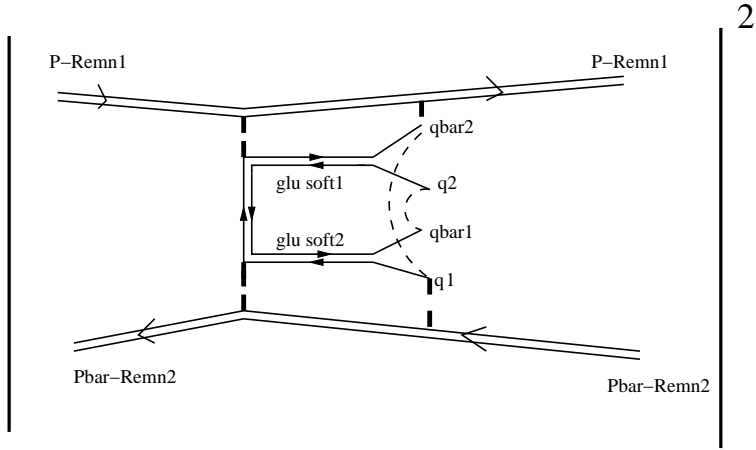,height=6cm}
\end{center}\caption{\footnotesize The soft collision between the two soft gluons, with dashed solid lines indicating the severed color connections between the remnants and the outgoing gluons, forming two clusters, $q_{1}\overline{q_{2}}$ and $q_{2}\overline{q_{2}}$.}
\end{figure}
\noindent

Because we wanted to keep the implementation of our model as simple as possible we have decided to sever the colour connections between the remnants and the outgoing soft gluons (see fig.2, dashed solid lines show severed colour connections). We expect this to be a reasonable approximation since the largest part of the phase space, in the gap between the two remnants, will be filled by final state hadrons produced by the colour field stretched in between the two outgoing soft gluons (note that in both figs 1 and 2, two colour lines stretch across the central region). 
 In the hadronisation phase the two outgoing soft gluons form two clusters ($q_{1}\overline{q}_{2}$, $q_{2}\overline{q}_{1}$) (see fig.2) which are decayed into final state hadrons (using HERWIG's cluster hadronisation model \cite{Marchesini:1992ch}).

\subsection{Implementation of the soft process}

Following fig.2, for each soft remnant-remnant interaction we generate two soft effective gluons, on mass shell, with previously tuned masses $m_{g}$ = 0.75 GeV used in the HERWIG hadronisation model \cite{Marchesini:1992ch}. For each soft interaction we will go through the following chain of events; first the maximum fractional longitudinal momentum allowed for each soft gluon ($x_{glumax \; 1,2}$) is determined using the light-cone definition of the longitudinal momentum fraction $\it{x}$ of the two remnants, 
\begin{equation}
x_{glumax \; 1,2} = x_{remn \; 1,2} = \frac{E_{remn1,2}+{P_{z_{remn1,2}}}}{E_{p,\overline{p}}+{P_{z_{p,\overline{p}}}}}.
\end{equation}               
The longitudinal momentum fractions for each soft gluon is then sampled from an $f(x)^{sea \; partons}$ = $\frac{1}{x}$ distribution (which we consider reasonable since we expect the effective gluons to be Regge-like) between some minimum value $x_{min}$ (the cutoff) and the maximum value, $x_{glumax \; 1,2}$, allowed by (32). Once their fractional longitudinal momentum has been determined, a transverse momentum for each soft gluon is then sampled from a Gaussian distribution for values of $p_{t}$, 0$\leq$$p_{t}$$\leq$$p_{tmin}$, according to
\begin{equation}
\frac{dN_{soft \; gluons}}{d^{2}p_{t}} = D \exp{(-\beta p_{t}^{2})}, 0 \leq p_{t} \leq p_{tmin}.
\end{equation}                                                                 
Before sampling the transverse momentum of each soft gluon, we will have to determine the slope $\beta$ of the particle $p_{t}$ distribution in the central region and the normalisation constant $D$ in (33). We impose then another condition on the $p_{t}$ distribution, namely, that the transverse momentum distribution in $p_{t}$ of soft and hard gluons should be continuous at the $p_{t}$ cutoff (34) (the same procedure was implemented in \cite{ralph:engl}),  
\begin{equation}             
\frac{dN_{soft \; gluons}}{d^{2}p_{t}}|_{p_{t} = p_{tmin}} = \frac{dN_{hard \; gluons}}{d^{2}p_{t}}|_{p_{t} = p_{tmin}}.
\end{equation}
We have then to solve two equations with two unknowns. The first condition is that the number of soft gluons should correspond to the soft cross section $\sigma_{SOFT}^{inc}(s_{p\overline{p}})$:

\begin{equation}
\int_{0}^{p_{tmin}}d^{2}p_{t} D \exp{(-\beta p_{t}^{2})} = \sigma_{SOFT}^{inc}(s_{p\overline{p}}),
\end{equation}      
the second condition is that of the smooth transition between the soft and hard perturbative gluons' transverse momentum distributions at the value of $p_{t}$ cutoff (34), re-expressed as:

\begin{equation}
D \exp{(-\beta p_{tmin}^{2})} = \frac{d\sigma(s_{p\overline{p}},p_{tmin})}{dp_{tmin}^2},
\end{equation}
where ${\frac{d\sigma(s_{p\overline{p}},p_{tmin})}{dp_{tmin}^2}}$ is the full differential parton-parton hard cross section calculated at the value of ${p_{t} = p_{tmin}}$. 

In order to determine the four momentum of each outgoing remnant and each soft gluon, two additional conditions need to be satisfied simultaneously; that of the total energy-momentum conservation between the initial beam-beam remnants (remn$_{1,2}$) and final (remn$_{1,2}$ + soft gluons) states and that the outgoing soft gluons and remnants are on mass shell, with each remnant having the same mass before and after the soft interaction.



Once all four momenta of the outgoing remnants and soft gluons have been determined a new maximum longitudinal fractional momentum ${x_{glumax \; 1,2}}$ is calculated, according to (32), for the next pair of soft gluons. The same chain of events, as described above, is then iterated until all soft interactions have been generated. If energy-momentum conservation is violated the scattering is vetoed.

\section{Comparison between the eikonal model and data from the CDF collaboration}
We will compare our model predictions with the experimental data coming from the study of the behaviour of the underlying event in hard scattering proton-antiproton collisions at 1.8 TeV. In ref.\cite{Affolder:2002zg} the underlying event in hard proton-antiproton collisions is defined as all the ambient interactions surrounding the hard scattered jets. These ambient interactions include beam-beam remnants, initial and final-state radiation and multiple hard interactions, if they do occur. The beam-beam remnants are just what is left over after a parton has been taken out from each of the initial incoming hadrons. The jets in a hard interaction are produced by streams of outgoing hadrons with large transverse momentum, from partons that have undergone a hard two-to-two scattering. In interactions where a hard two-to-two subprocess occurred, jets of hadrons are produced approximately back to back in azimuthal angle. This simple jet structure allows for the study of the particle distribution in azimuthal angle $\phi$. There are three regions of study separated in $\Delta \phi$ from the leading highest $p_{t}$ jet (see fig.3). The ``toward'' region containing the leading charged particle jets, the ``away'' region containing particles produced opposite in $\phi$ from the leading jet and the ``transverse'' region perpendicular to the plane of the hard two-to-two scattering which is the most sensitive region to the underlying event.         

\begin{figure}[h]
\begin{center}
\epsfig{figure=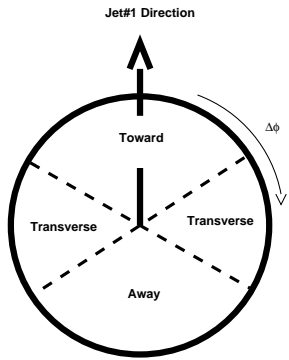,height=5cm}
\end{center}\caption{\footnotesize Toward, away and transverse regions from the leading in $p_{t}$ jet (see ref. \cite{Affolder:2002zg}).The angle $\Delta$$\phi$ = $\phi$ - ${\phi_{jet\#1}}$ is the relative azimuthal angle between charged particles and the direction of jet$\#1$. The three regions are as follows; the toward region defined over the range  $\mid$$\Delta$$\phi$$\mid$ $<$ ${60^{\circ}}$ (this region includes the particles of the leading jet$\#1$), the away region  defined over the range  $\mid$$\Delta$$\phi$$\mid$ $>$ ${120^{\circ}}$ and the transverse region defined over the range ${60^{\circ}}$ $<$ $\mid$$\Delta$$\phi$$\mid$ $<$ ${120^{\circ}}$. Each region, toward, transverse, and away covers the same range $\mid$$\Delta$$\eta$$\mid$$\times$$\mid$$\Delta$$\phi$$\mid$=2$\times$${120^{\circ}}$.}
\end{figure}

The simple jet algorithm used in \cite{Affolder:2002zg} builds jets from charged particles with $p_{t}$ $>$ 0.5 GeV and $\mid$$\eta$$\mid$ $<$ 1 and is restricted to particle jets with transverse momentum less than 50 GeV. The charged particles form jets of radius $\it{R}$ = 0.7 in $\eta$-$\phi$ space which contain particles from both underlying event and hard scattering. Every charged particle is assigned to a jet with a possibility that some jets might consist of just one charged particle. The transverse momentum of the jet is defined as the scalar $p_{t}$ sum of all particles within the jet. Once all jets have been produced the one with the highest $p_{t}$(jet$\#1$) is defined to be the leading jet. The experimental data are uncorrected and the theoretical simulated data are corrected for the track finding efficiency by removing, on average 8$\%$, of the charged particles.  

\subsection{Results}
A comparison between simulated (corrected) and experimental data is made and presented in figs 4-7, with $p_{tmin}$ = 2.5 GeV. The agreement seen in figs.6 and 7 is particularly important since the transverse region is the most sensitive to the underlying event activity. 
In figs.8 and 9 we compare our predictions for the transverse region with those of the existing models (without including Pythia), HERWIG + Underlying Event model \cite{Marchesini:1992ch}, HERWIG + Multiparton Hard model \cite{Butterworth:1996zw}. In HERWIG's default Underlying Event model, the activity is enhanced in the transverse region by adding to its hard part a Minimum Bias Event \cite{Marchesini:1988hj}, based on the parametrisation of the UA5 ${p\overline{p}}$ minimum-bias data. As shown in figs.8 and 9 and in \cite{Affolder:2002zg} this model fails to produce enough activity for both the average number of charged particles and the average scalar $p_{t}$ sum and in addition has a wrong (too steep) $p_{t}$ dependence. Overall our eikonal model is in better agreement with the experimental data than either of the two underlying event models. This result is even more significant since we did not fit to the experimental data but rather, make predictions based on the value of the total cross section used as phenomenological input to our model.     

\subsection{The invariance of the model to $\bf{p_{tmin}}$}
In the multiparton scattering model \cite{Butterworth:1996zw}, the parameter $p_{tmin}$ is a cutoff scale and plays a crucial role in determining the prediction of the model. In our eikonal model it is rather a matching scale between the hard and soft parts of the model. If the matching works perfectly, the results should be $p_{tmin}$ independent.

To test the invariance of the eikonal model to its $p_{tmin}$ parameter we simulate two sets of data, with two different values of $p_{tmin}$ = 2 GeV and 3 GeV in addition to the value of 2.5 GeV, which we have already shown. The values of inclusive cross sections with the average number of soft and hard scatters are presented in table.1.

\begin{table}[h]
\begin{center}
\begin{tabular}{|c|c|c|c|c|}\hline 
 $\it{p_{tmin}}$(GeV) &$\sigma_{SOFT}^{inc}(s_{p\overline{p}})$(mb) &$\sigma_{H}^{inc}(s_{p\overline{p}})$(mb) & $\langle n_{soft} \rangle$  &$\langle n_{hard} \rangle$ \\\hline
2.0    &39.7       &99.2   &0.7  &1.7  \\ 
2.5    &85.6       &51.3   &1.5  &0.9  \\
3.0    &109.7      &28.7   &1.9  &0.5   \\\hline      
\end{tabular}
\end{center}
\caption{Values of inclusive cross sections and average numbers of hard and soft scatters.}
\end{table}

We show the results in fig.10. We see that as expected the results in the multiparton model are strongly $p_{tmin}$ dependent, with smaller values of $p_{tmin}$ producing more activity. However, even the smallest value of $p_{tmin}$ is well below the data and no value of $p_{tmin}$ for which the hard part of HERWIG is reliable gives a good description of the data (although it is worth noting that with $p_{tmin}$ fixed at 2.5 GeV it is possible to fit the data by increasing the proton radius by a factor of about 1.7 \cite{I.Borozan}. Since this model does not provide any explanation of how the radius should vary as a function of center-of-mass energy, or for different collision types, we do not consider it to be useful phenomenologically).
In contrast, our eikonal model has a smaller $p_{tmin}$ dependence, with the additional soft scatters providing extra activity to compensate for that lost from the hard scatters as $p_{tmin}$ increases. However, we see that they actually overcompensate, with decreasing $p_{tmin}$ leading to decreasing activity. This residual $p_{tmin}$ dependence indicates that the matching of the soft and the hard parts of the model is still not perfect and can perhaps be improved with further refinements of the model. Nevertheless, we are satisfied that with such a simple, physically motivated and parameter free model, we have provided a significant improvement to the $p_{tmin}$ dependence and the description of data.

\clearpage

\begin{figure}
\begin{center}
\epsfig{figure=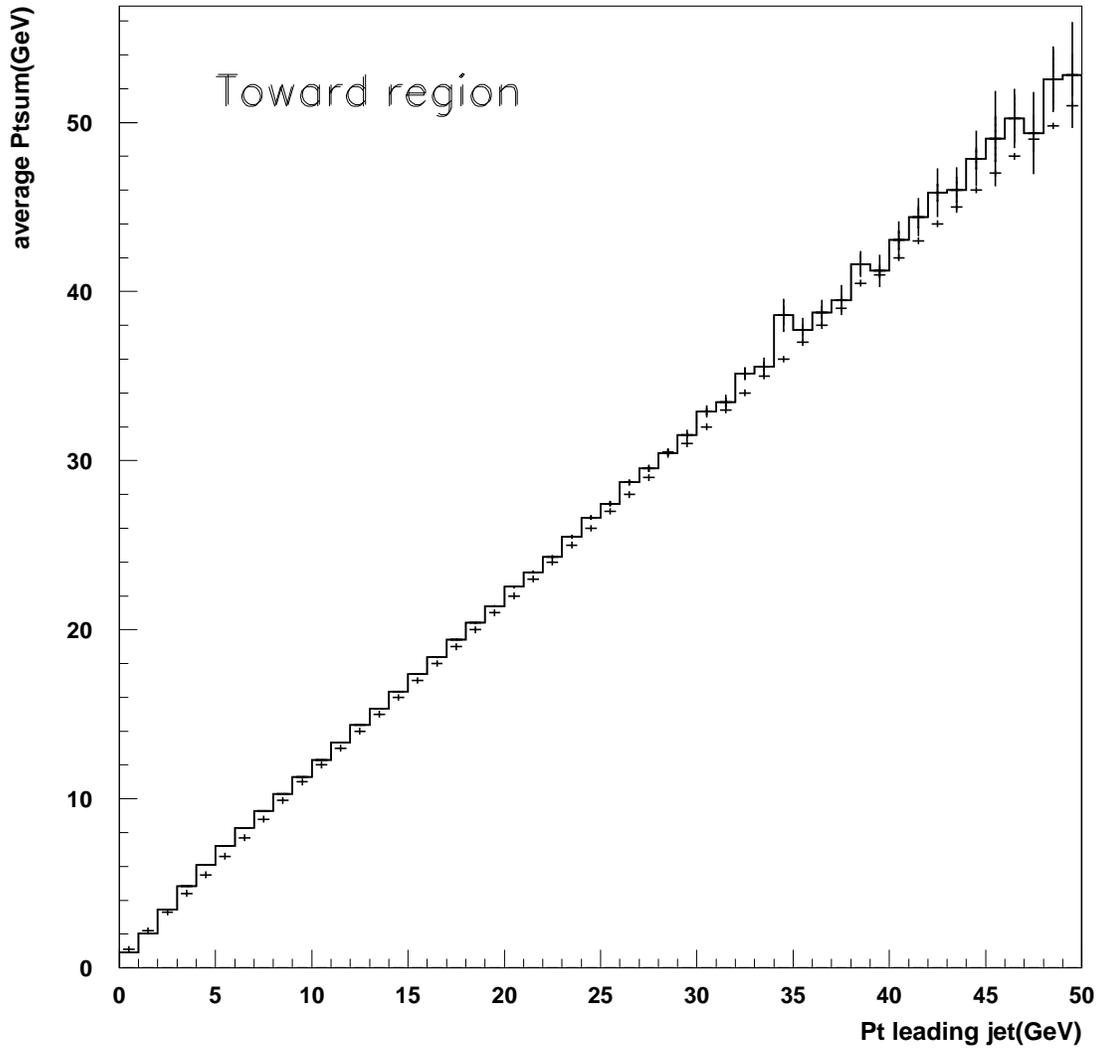,height=16cm}
\end{center}\caption{\footnotesize The average $P_t$ sum of charged particles as a function of $P_t$ (leading charged jet) in the toward region. HERWIG + Eikonal model solid line simulated data, experimental data solid circles.}
\end{figure}

\clearpage

\begin{figure}
\begin{center}
\epsfig{figure=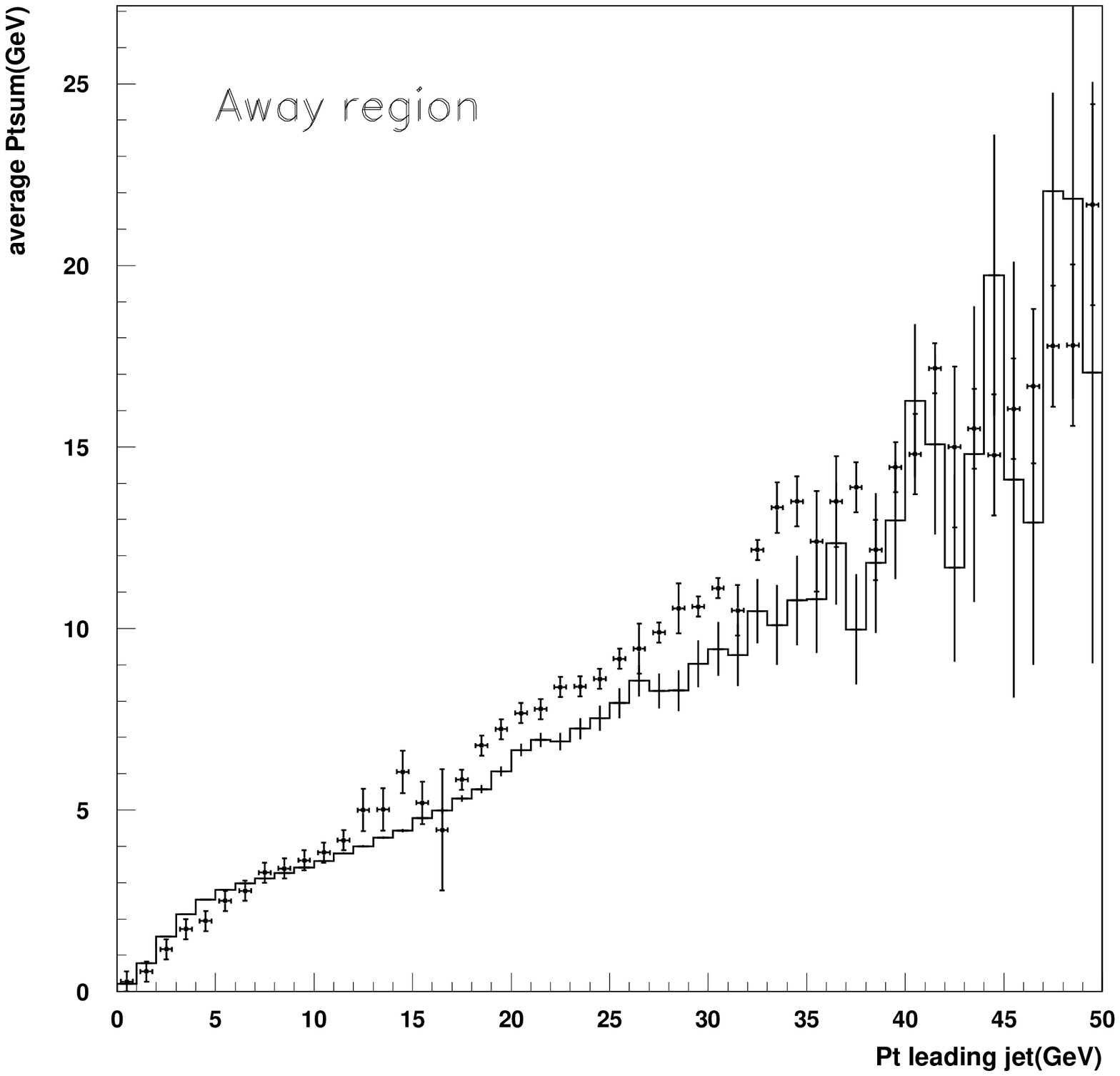,height=16cm}
\end{center}\caption{\footnotesize The average $P_t$ sum of charged particles  as a function of $P_t$ (leading charged jet) in the away region. HERWIG + Eikonal model solid line simulated data, experimental data solid circles.}
\end{figure}

\clearpage

\begin{figure}
\begin{center}
\epsfig{figure=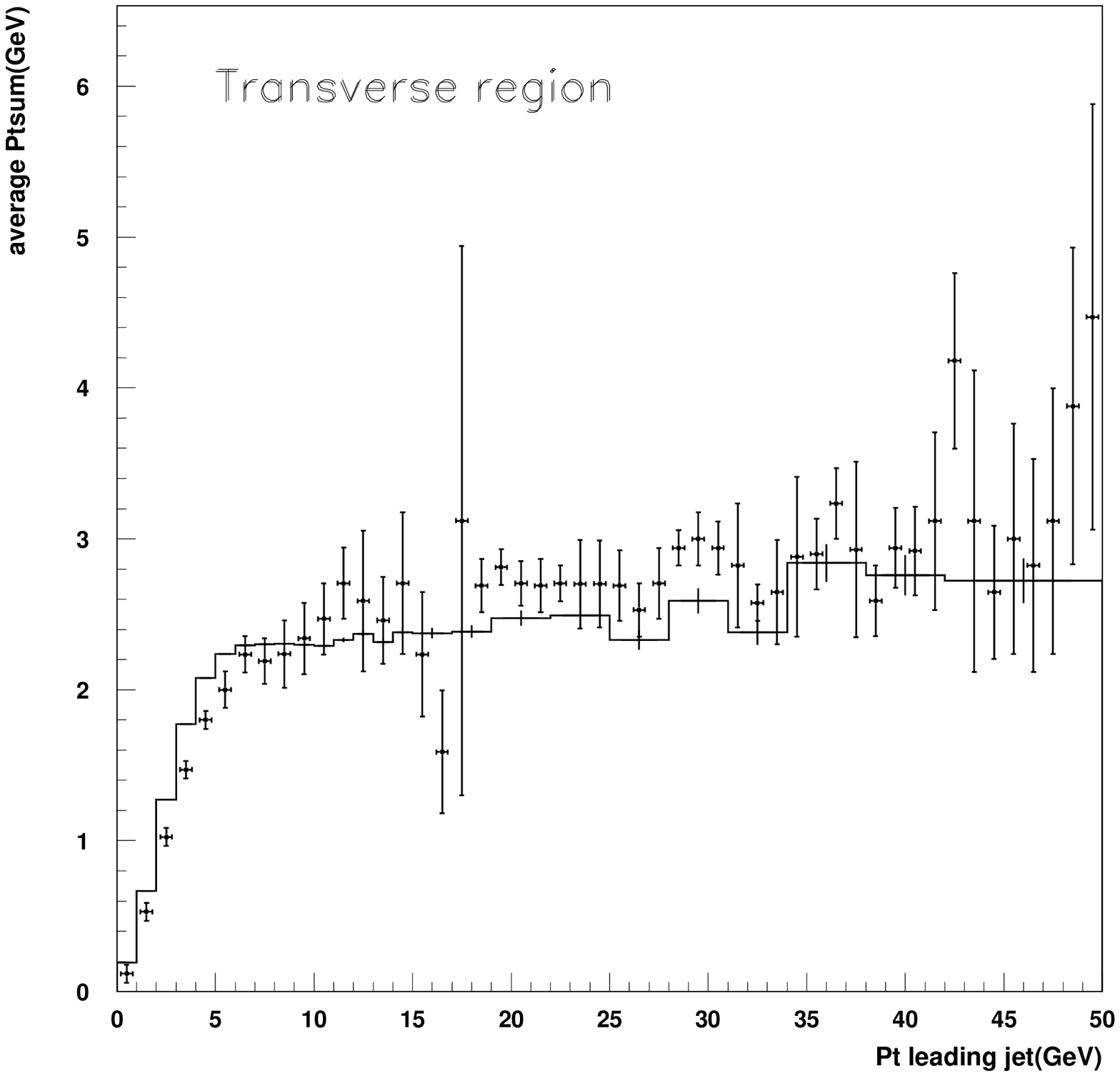,height=16cm}
\end{center}\caption{\footnotesize The average $P_t$ sum of charged particles  as a function of $P_t$ (leading charged jet) in the transverse region. HERWIG + Eikonal model solid line simulated data, experimental data solid circles.} 
\end{figure}

\clearpage

\begin{figure}
\begin{center}
\epsfig{figure=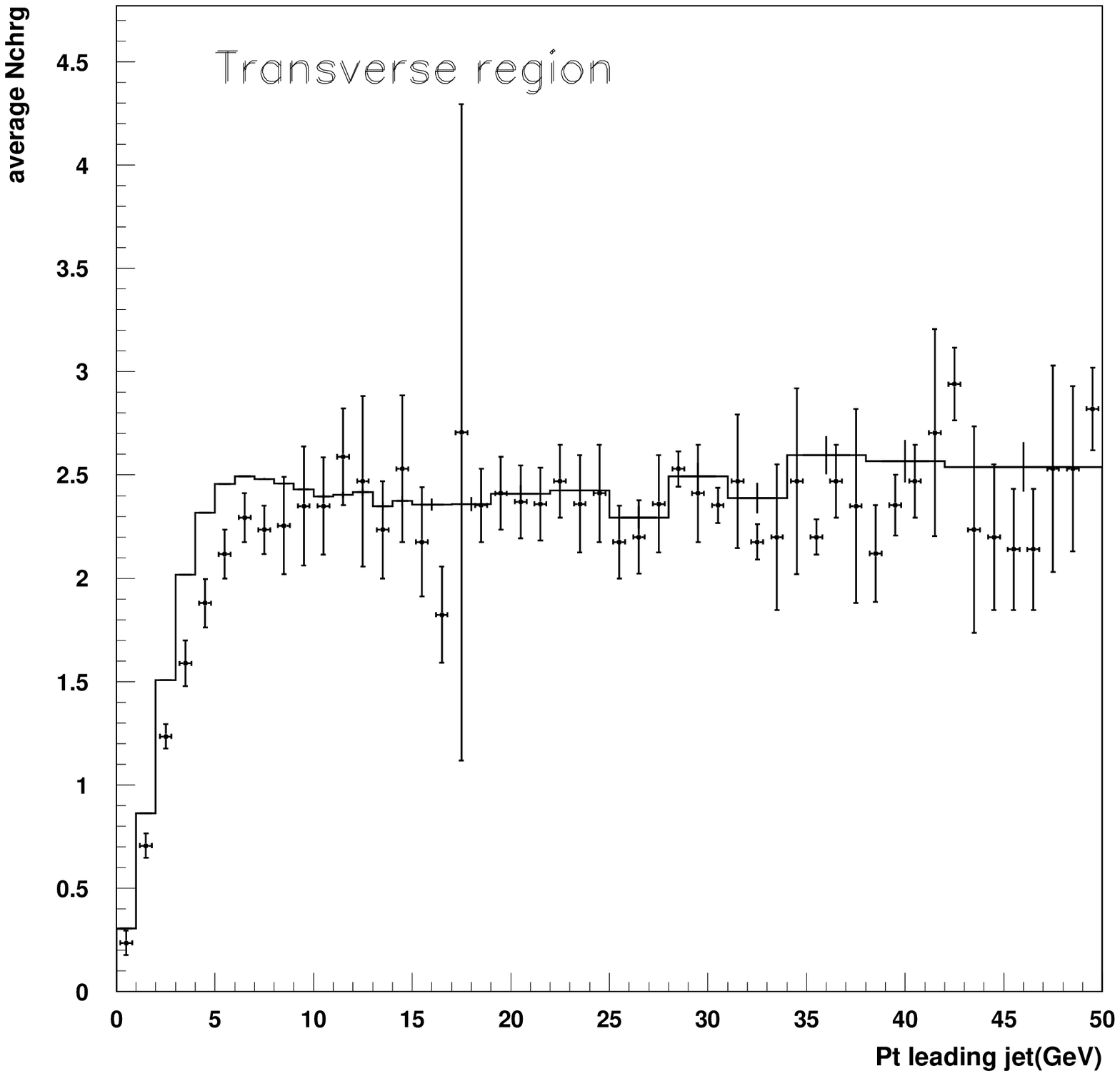,height=16cm}
\end{center}\caption{\footnotesize The average number of charged particles  as a function of $P_t$ (leading charged jet) in the transverse region. HERWIG + Eikonal model solid line simulated data, experimental data solid circles.}
\end{figure}

\clearpage

\begin{figure}
\begin{center}
\epsfig{figure=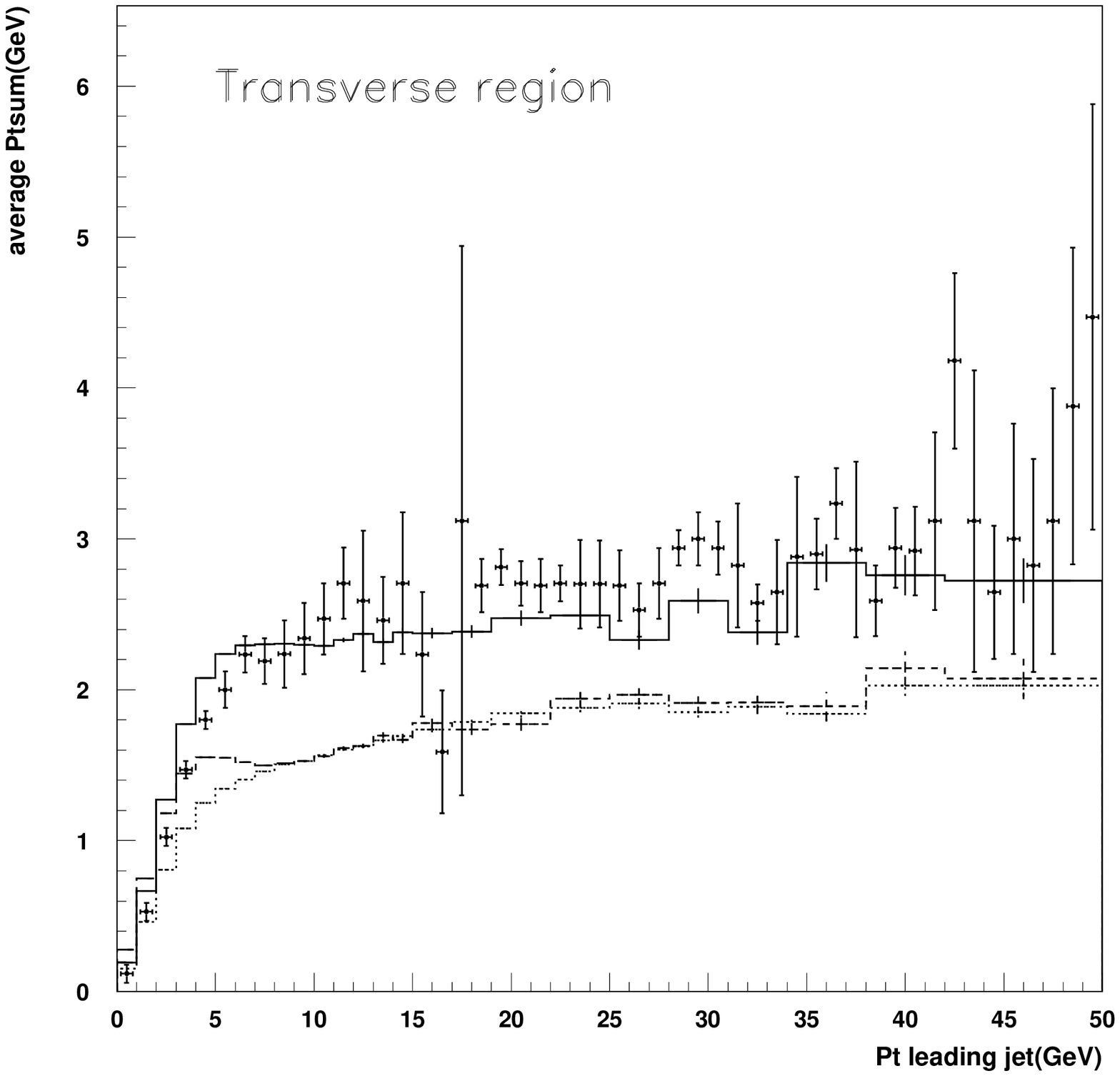,height=16cm}
\end{center}\caption{\footnotesize The average $P_t$ sum of charged particles  as a function of $P_t$ (leading charged jet) in the transverse region. HERWIG + Eikonal model  (solid line), HERWIG + Underlying Event model (solid dashed), HERWIG + Multiparton Hard model (dotted), experimental data solid circles.}
\end{figure}

\clearpage

\begin{figure}
\begin{center}
\epsfig{figure=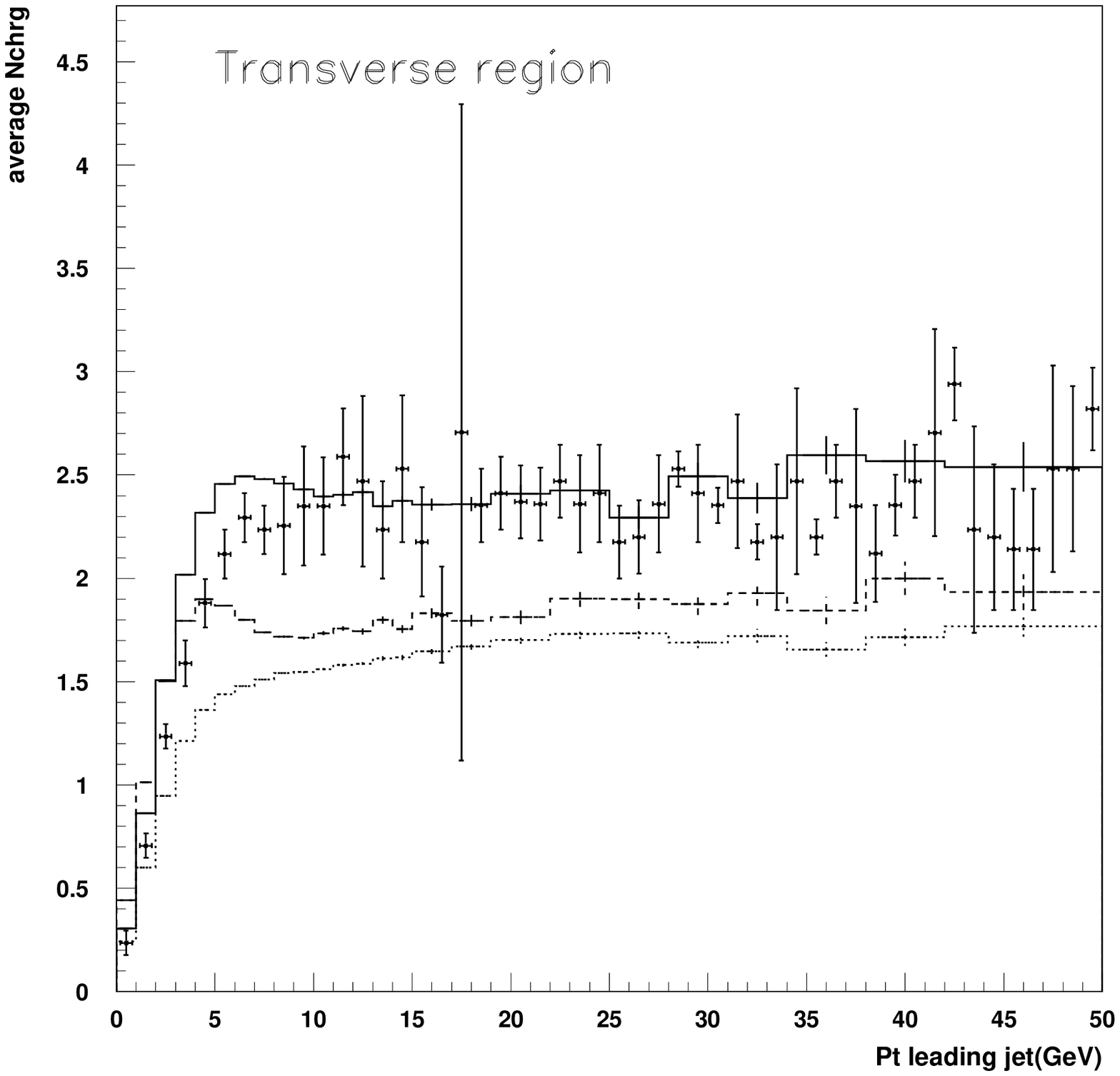,height=16cm}
\end{center}\caption{\footnotesize The average number of charged particles as a function of $P_t$ (leading charged jet) in the transverse region. HERWIG + Eikonal model (solid line), HERWIG + Underlying Event model (solid dashed), HERWIG + Multiparton Hard model (dotted), experimental data solid circles.} 
\end{figure}

\clearpage

\begin{figure}
\begin{center}
\epsfig{figure=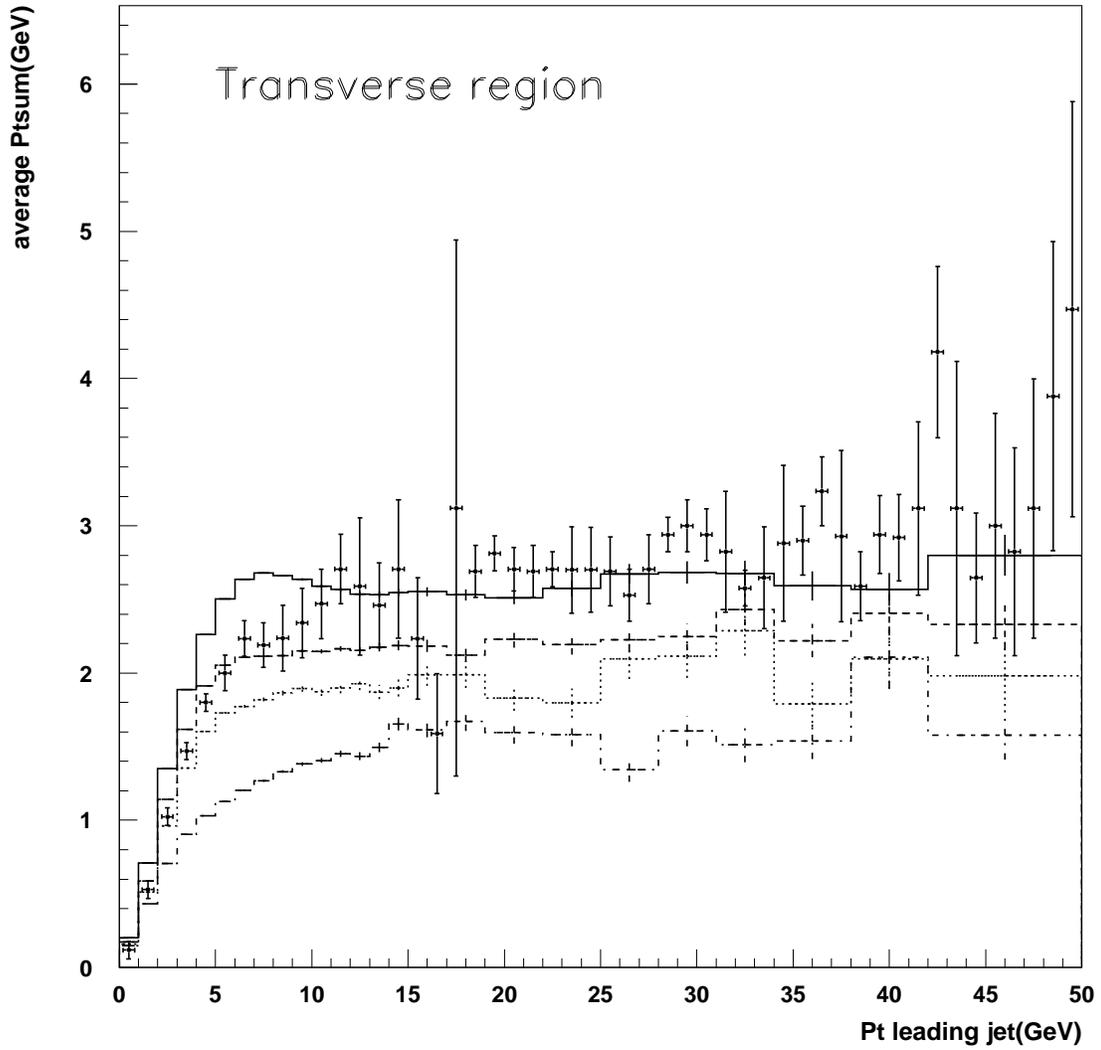,height=16cm}
\end{center}\caption{\footnotesize The average $P_t$ sum of charged particles  as a function of $P_t$ (leading charged jet) in the transverse region. HERWIG + Eikonal Model for the two sets of ${p_{tmin}}$=3.0 GeV (solid line) and 2.0 GeV (dashed), HERWIG + Multiparton Hard Model ${p_{tmin}}$=2.0 GeV (dotted),  ${p_{tmin}}$=3.0 GeV (dott dashed), experimental data points.}  
\end{figure}

\clearpage

\section{Discussion and Conclusion}
In this paper we have proposed a Monte Carlo model, based on a simple eikonal model, for multiparticle production in hard proton-antiproton collisions. The model contains two parts, the hard part, which was already implemented \cite{Butterworth:1996zw} (running in conjunction with HERWIG), to which we have added a soft part, which uses showering and hadronisation models from HERWIG and allows us to extend our simulations to the non-perturbative soft region (i.e. particles with $0 \leq p_{t} \leq p_{tmin}$). One of our goals was to produce a simple model (with a minimum number of parameters) which would be reasonably independent of the value chosen for the $p_{tmin}$. Furthermore, the eikonal multiparticle approach to hadron-hardon scattering is particularly interesting since it prevents the calculated cross sections from violating unitarity at high energies. 
By fixing one parameter in our model, namely the total cross section, we provide a good description of the measured data from \cite{Affolder:2002zg} with a reasonably small ${p_{tmin}}$ dependence. We have also shown that our model gives a better prediction than either the HERWIG + Underlying Event \cite{Marchesini:1992ch} or HERWIG + Hard Multiparton \cite{Butterworth:1996zw} models. 

Our ultimate goal is to determine how much of the underlying event activity in the transverse region is due to the perturbative physics (by including NLO calculations for example) and how much is due to non-perturbative physics, and to make more reliable predictions of the underlying event. We are planning to test the model with the lower energy UA1 collaboration data, HERA data, higher ${p_t}$ CDF data and give a prediction for the underlying event activity at the LHC.\\

\noindent
{\Large{\bf{Acknowledgements}}} \\
I.Borozan would like to express his gratitude to CLRC for funding his studies and to Manchester Theoretical Group for their hospitality. For useful discussions we thank Jon Butterworth and Jeff Forshaw.

\end{document}